\documentclass[useAMS,usenatbib]{mn2e}
\usepackage{subfigure}
\usepackage{graphicx}
\usepackage{amsmath}
\usepackage{verbatim}
\usepackage{multirow}
\usepackage{upgreek}
\usepackage{textcomp}
\usepackage{rotating}
\usepackage{color}
\usepackage{subdepth}
\pdfoutput=1
\title[Zinc abundances of planetary nebulae]{Zinc abundances of planetary nebulae}
\author[C. L. Smith, A. A. Zijlstra and H. L. Dinerstein]{C. L. Smith $^{1}$\thanks{E-mail: Christina.Smith-3@postgrad.manchester.ac.uk}, A. A. Zijlstra$^{1}$ and H. L. Dinerstein$^{2}$ \\ 
$^{1}$ Jodrell Bank Centre for Astrophysics, University of Manchester, Manchester, M13 9PL, UK \\ $^{2}$ Department of Astronomy and McDonald Observatory, University of Texas, Austin, TX 78712-1083}

\begin{document}

\date{XXX}

\pagerange{\pageref{firstpage}--\pageref{lastpage}} \pubyear{2013}

\maketitle

\label{firstpage}

\begin{abstract}
Zinc is a useful surrogate element for measuring Fe/H as, unlike iron, it is not depleted in the gas phase media. Zn/H and O/Zn ratios have been derived using the [Zn IV] emission line at 3.625$\upmu$m for a sample of nine Galactic planetary nebulae, seven of which are based upon new observations using the VLT.   Based on photoionization models, O/O$^{++}$ is the most reliable ionisation correction factor for zinc that can readily be determined from optical emission lines, with an estimated accuracy of 10\% or better for all targets in our sample.   The majority of the sample is found to be sub-solar in [Zn/H]. [O/Zn] in half of the sample is found to be consistent with Solar within uncertainties, whereas the remaining half are enhanced in [O/Zn]. [Zn/H] and [O/Zn] as functions of Galactocentric distance have been investigated  and there is little evidence to support a trend in either case. 

\end{abstract}


\section{Introduction}

Iron has been shown to be heavily depleted from the gaseous form in ionised media through both optical \citep{Delgado2009, Shields1975} and infrared observations \citep{Likkel2006}. In planetary nebulae it can be depleted by  more than  90\% \citep{Delgado2009}. As shown in \citet{Sterling2005}, depletion may be not be uniform throughout a planetary nebula.  Iron depletion is also seen in many environments in the interstellar medium (ISM) \citep{McDonald2010, Savage1996}. Therefore deriving iron abundances of planetary nebulae through direct observations of iron will not result in an accurate elemental abundance for the precursor star. 

Zinc, of which there are five different natural isotopes, exists most abundantly as $^{64}$Zn and is predominantly created in two processes: the alpha rich freeze-out of supernovae and the s-process \citep{Clayton2003, Woosley1995}. The condensation temperature of zinc is 684 K which is significantly lower than that of the other iron-peak elements which range from 1170 K for Cu to 1600 K for Ti, with iron itself having a condensation temperature of 1337 K \citep{Lodders1998}. This low condensation temperature means it is significantly less likely for  zinc to condense into solids in the progenitor star's envelope than it is for iron. 

Zinc abundances and depletion in the interstellar medium have been widely studied over the past several decades (e.g. \citealp{Morton1975}). In warm, low-density regions of the interstellar medium, zinc has been shown to be depleted by 0.0-0.2 dex \citep{Savage1996, Welty1999}.

The abundance of zinc over a wide range of metallicities within the Milky Way was studied by \citet{Sneden1991} by examining a large sample of field and globular cluster stars. The Zn/M abundance was shown to be constant over a wide range of M/H, where M/H is defined as the average metallicity of the star derived from observations of Fe and Ni. There have been some recent studies which find some discrepancies between Zn and Fe abundances in Galactic stars: \citet{Prochaska2000}, for example, find a mean enhancement of [Zn/Fe] to be  0.093 $\pm$ 0.025 from a sample of ten thick disk stars and \citet{Chen2004} find [Zn/Fe] to be unenhanced in a sample of five alpha-poor halo stars but enhanced by 0.15 dex in a sample of ten thick disk stars. However, these are based upon much smaller samples than that studied by \citet{Sneden1991}.  \citet{Saito2009}, report measurements of [Zn/Fe] of a sample of 35 stars to be Solar to within uncertainties over the range $-2<\text{[Fe/H]}<0$. \citet{Saito2009} also combine their measurements with that of 399 literature stars with the resulting sample covering the range $-4.5<\text{[Fe/H]}<0.5$. [Zn/Fe] was found to be consistent with solar in the ranges $-2<\text{[Fe/H]}<+0.5$ to within the measured standard deviation.

 In planetary nebulae,  the 3.625 $\upmu$m [Zn IV] line is well suited to abundance derivation as it lies between two hydrogen Humphreys series emission lines (3.646 $\upmu$m and 3.607 $\upmu$m) and less than 0.125 $\upmu$m away from the much stronger Pfund series emission line at 3.741 $\upmu$m. Zn$^{3+}$, in contrast to most of the abundant Fe ions in nebulae, has a sparse energy level structure in its ground configuration, making the 3.625 $\upmu$m line relatively strong.  The 3.625 $\upmu$m line is the only emission line of zinc observed in the infrared in planetary nebulae.   

The 3.625 $\upmu$m [Zn IV] line was first identified by \citet{Dinerstein2001} in spectra taken using CGS4 on the UK Infrared Telescope (UKIRT) of NGC 7027 and IC 5117. They find the central wavelength, after corrections for the Earth's orbital motion, to be within uncertainties of that predicted for the Zn$^{3+}$ $^2\text{D}_{3/2}-^2\text{D}_{5/2}$ transition. They consider alternative identities of the line from nearby transitions of H$_2$ or Cu$^{4+}$. In the case of H$_2$, this is ruled out on the basis of the observed wavelength of the line. The $^4\text{F}_{5/2}-^4\text{F}_{9/2}$ Cu$^{4+}$ transition was deemed less likely than the $^2\text{D}_{3/2}-^2\text{D}_{5/2}$ transition of Zn$^{3+}$ due to the lower elemental abundance, stronger depletion and higher ionisation state of the former species. 

\citet{Dinerstein2001} presented the procedure for deriving $\rm{Zn}^{3+}/H^+$ from their measurements. However, since the collision strength of the Zn transition was not known at the time, their abundance estimates were necessarily expressed in terms of the value of this parameter (their equation 4). After correcting for the presence of other ions of zinc, elemental Zn/H can be determined. Dinerstein \& Gaballe suggested that such Zn/H values are excellent tracers of the total Fe/H abundances and that [Zn/H] can be used as a surrogate for [Fe/H], in view of the minimal depletion of Zn and the fact that Zn closely tracks Fe over a wide range in metallicity.

Small contributions of the elemental abundance of zinc are expected from nucleosynthesis in AGB stars. The models of \citet{Karakas2009} show this enhancement of Zn over the course of a star's time on the AGB is small: only the very low metallicity models show [Zn/Fe] $> 0.1$, and the majority of the remaining models show [Zn/Fe] $<0.05$. Thus the measured abundance of Zn in planetary nebulae will, in general, reflect the abundance of Zn of the progenitor star.

The ability to determine meaningful Fe/H values in nebulae is important because Fe/H is the most widely used index of metallicity in stars; therefore by determining undepleted Fe/H ratios, reliable metallicities of the precursor stars of the planetary nebulae can be determined. By examining the O/Zn ratio as a proxy for O/Fe, $\alpha$-element enhancement in these objects can be assessed. 

The remainder of this paper will detail measurements of zinc abundances of a sample of planetary nebulae using the 3.625 $\upmu$m emission line, and the extrapolated iron abundances.

\begin{table*}
\centering
\caption{Basic information and literature parameters for each source. Angular diameters are in arc seconds, where $^\text{opt}$ indicates values derived from optical observations and $^\text{rad}$ denotes values derived from radio observations \citep{Acker1992}. The first  seven   nebulae are those with new observations, and the remaining two are from \citet{Dinerstein2001}. References for optical spectra and  central star effective temperatures  are listed in column 11.}\label{basic_info}
{\tabcolsep=0.12cm
\begin{tabular}{c c c c c c c c c c c}
\renewcommand{\arraystretch}{1.5}
PNG & Name & RA & Dec & Ang. dia. & T$_{\text{star}}$ & T$_\text{e}$(O III) & N$_\text{e}$(S II) & O/H & O$^{++}$/O & Ref. \\
 & & & & (arcsec) & ($\times10^4$ K) & ($\times10^4$ K) & ($\times10^3$ cm$^{-3}$) & ($\times10^{-4}$) & \\
\hline\hline
$004.0-03.0$  & M 2-29 &  18 06 40.86 & $-$26 54 55.95 & 3.6$^\text{opt}$ & 7.6 & $1.9\pm{0.3}$  &  $3\pm{3}$ &  $0.3^{+0.1}_{-0.2}$  & $0.9^{+0.5}_{-0.8}$ & 1 ,2 \\
$006.1+08.3$  & M 1-20 & 17 28 57.61 & $-$19 15 53.94 & 1.9$^\text{rad}$& 7.9 & $1.01\pm{0.02}$ &  $6^{+3}_{-1}$  &  $3.4\pm{0.3}$   & $0.93\pm{0.13}$ & 3 ,2 \\
$006.4+02.0$  & M 1-31 & 17 52 41.44 & $-$22 21 57.00 & 7.0$^\text{rad}$& 5.8 & $0.76\pm{0.04}$  &  $8^{+4}_{-1}$  &  $7.7\pm{1.6}$  & $0.94\pm{0.28}$ & 4 ,2 \\
$006.8+04.1$ & M 3-15 & 17 45 31.71 & $-$20 58 01.57 & 4.2$^\text{opt}$& 7.9  &  $0.85\pm{0.02}$ &  $5^{+3}_{-1}$ &  $6.0\pm{0.7}$  & $0.96\pm{0.16}$ & 4 ,2 \\
$019.7+03.2$ & M 3-25 & 18 15 16.97 & $-$10 10 09.47 & 3.9$^\text{opt}$& 5.2 &  $1.09\pm{0.03}$  &  $14\pm{1}$ &  $3.9\pm{0.4}$  & $0.82\pm{0.11}$ & 5 ,6 \\
$040.4-03.1$ & K 3-30 & 19 16 27.70 & +05 13 19.40 & 3.3$^\text{rad}$ &  - & 1.0$^\text{a}$  &  10.0$^\text{a}$  &  3.9$^\text{a}$  &  0.80$^\text{a}$ & -  \\
$355.4-02.4$ & M 3-14 & 17 44 20.62 & $-$34 06 40.60 & 2.8$^\text{rad}$ & 7.9 &  $0.87\pm{0.04}$  &  $3.4\pm{0.2}$ &  $6.5\pm{0.1}$  & $0.73\pm{0.03}$ & 7 ,2 \\
\hline
$084.9-03.4$ & NGC 7027 & 21 07 01.59 & +42 14 10.18 & 14.0$^\text{opt}$ &  18.0 &  $1.25\pm{0.04}$  &  $13^{+10}_{-3}$ &  $3.9\pm{0.5}$  &  $0.67\pm{0.01}$ & 8 ,9 \\
$089.8-05.1$ & IC 5117 & 21 32 30.97 & +44 35 47.5 & 1.5$^\text{rad}$&  12.0 &  $1.25\pm{0.04}$ &  $16^{+11}_{-4}$ &  $2.7\pm{0.3}$  &  $0.94\pm{0.13}$ & 10 ,10 \\
\end{tabular}
}
{\begin{flushleft}
1: \citet{Exter2004};  2: \citet{Gesicki2007};   3: \citet{Wang2007}; 4: \citet{Gorny2009}; 5: \citet{Gorny2004};  6: \citet{Kondratyeva2003};  7: \citet{Cuisinier2000} ; 8: \citet{Zhang2005};  9: \citet{Pottasch2010};   10: \citet{Hyung2001}.
\vspace{0.3cm} 

a: adopted value
\end{flushleft}}
\end{table*}

\section{Observations and data reduction}

Nine planetary nebulae, five of which are Galactic bulge nebulae,  were observed using ISAAC (Infra-red Spectrometer And Array Camera, \citet{Moorwood1998}) on UT3 at ESO Paranal, Chile between 28/05/2012 and 30/05/2012. These nebulae were selected from the Catalogue of Galactic Planetary Nebulae \citep{Acker1992} and were chosen based upon their galactic position, size of the nebula ($<$ 5" in diameter), brightness, excitation and the intensity of the O III emission lines. 

Our program was designed to target planetary nebulae belonging to the Galactic bulge. However, four nebulae not belonging to the bulge, namely PNG 019.7+03.2, PNG 040.4--03.1, PNG 023.0+04.3 and PNG 049.3+88.1 were observed as back-up targets when high winds caused pointing restrictions during part of the run. The last two objects yielded no detections or useful limits, and are not discussed further. Of the nine nebulae observed, seven were detected in the 3.625 $\upmu$m line including five bulge sources.

The observations used the Long Wavelength Spectrometer using the jiggle-nod method, with on-source exposure times of 30-60 minutes. The slit length was 120 arcsec, the slit width used was 1 arcsec. The wavelength coverage was 3.55-3.80 $\upmu$m and the resulting resolution was $R=2000$. Nodding was done along the slit, with nods of 15 to 30 arcsec depending upon the source being observed. This results in the source being in the field of view at all times during the observation, reducing the required observing time whilst still allowing background subtraction to be carried out effectively. The slit was aligned through the brightest portion of the nebulae and in most cases included more than 50\% of the object, due to their small angular size.  The position angle of the slit was 0$^{\circ}$ for all observations. 

The detector spectral coverage in this wavelength domain and at this resolution is 0.255 $\upmu$m, allowing the [Zn IV] emission line to be observed concurrently with five lines from the Humphreys series of hydrogen (n=21-6 to n=17-6) and one Pfund series hydrogen emission line (n=8-5, henceforth H$_{8-5}$).

The data were reduced using the jiggle-nod recipes provided by the ESO common pipeline library, including flat fielding. The wavelength calibrations were done using arc lamp lines rather than sky lines, as recommended in the ISAAC user manual, due to there being no usable sky emission lines within this wavelength range.

The data were relative flux calibrated using the telluric standard stars observed with each planetary nebula observation. These telluric standard stars are almost exclusively B-type stars taken from the Hipparcos Catalogue\footnote{ESA, 1997, The Hipparcos and Tycho Catalogues, ESA SP-1200}, and are listed in Table \ref{tellurics}. Their stellar types are well documented and, from these types, an effective temperature of each star could be estimated. In this spectral region,  these   stars are well described by blackbody distributions calculated from their effective temperatures. The correction function to apply across the spectrum was calculated by dividing the blackbody function of each of the standard stars by the observed telluric stellar spectrum, after the removal of stellar spectral features, and normalising.  The difference between the calibration using the B-type stars and the K-type stars is that extra features exist in the spectra of the B-type stars that must be removed before the correction function is calculated.  The observed planetary nebulae data were then multiplied by the normalised functions to produce relative flux calibrated data. One nebula was observed twice and its two resulting relative flux calibrated spectra were averaged.  There was no significant difference in the intensities of the two spectra. We estimate the relative flux calibrations to be accurate to 5-10\%. 

Absolute flux calibrations are not required for the analysis done here because both the hydrogen and zinc lines are contained within the same spectrum and only the ratio of the two is required to derive the zinc-to-hydrogen abundance. 

The spectra of sources with detections of the [Zn IV] line and several H I lines are shown in Fig. \ref{spectra}. Two spectral ranges are shown for each source so that the intensity of both the H$_{8-5}$ and [Zn IV] lines can be clearly seen. 

In order to measure the integrated line intensities for abundance calculations, the spectra were first smoothed, then a straight-line background level was fitted and the integrated intensities of lines measured, after correcting for obvious noise spikes. The abundance calculations were then carried out as described in Section \ref{abd_der}.

\begin{figure*}
\centering
\subfigure{\includegraphics[trim=0cm 12cm 0cm 3cm, clip=true,scale=0.4]{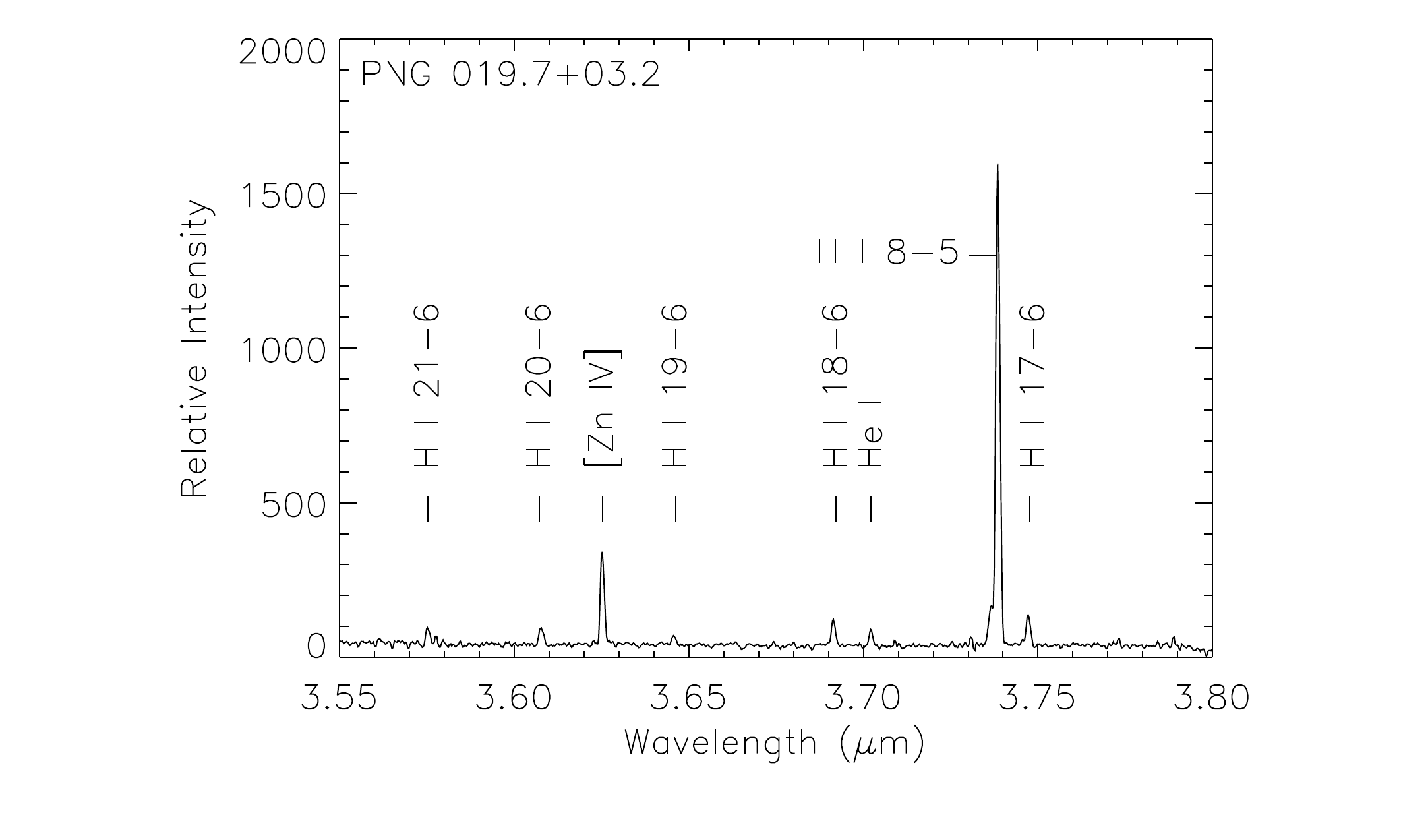}}
\subfigure{\includegraphics[trim=0cm 12cm 0cm 3cm, clip=true,scale=0.4]{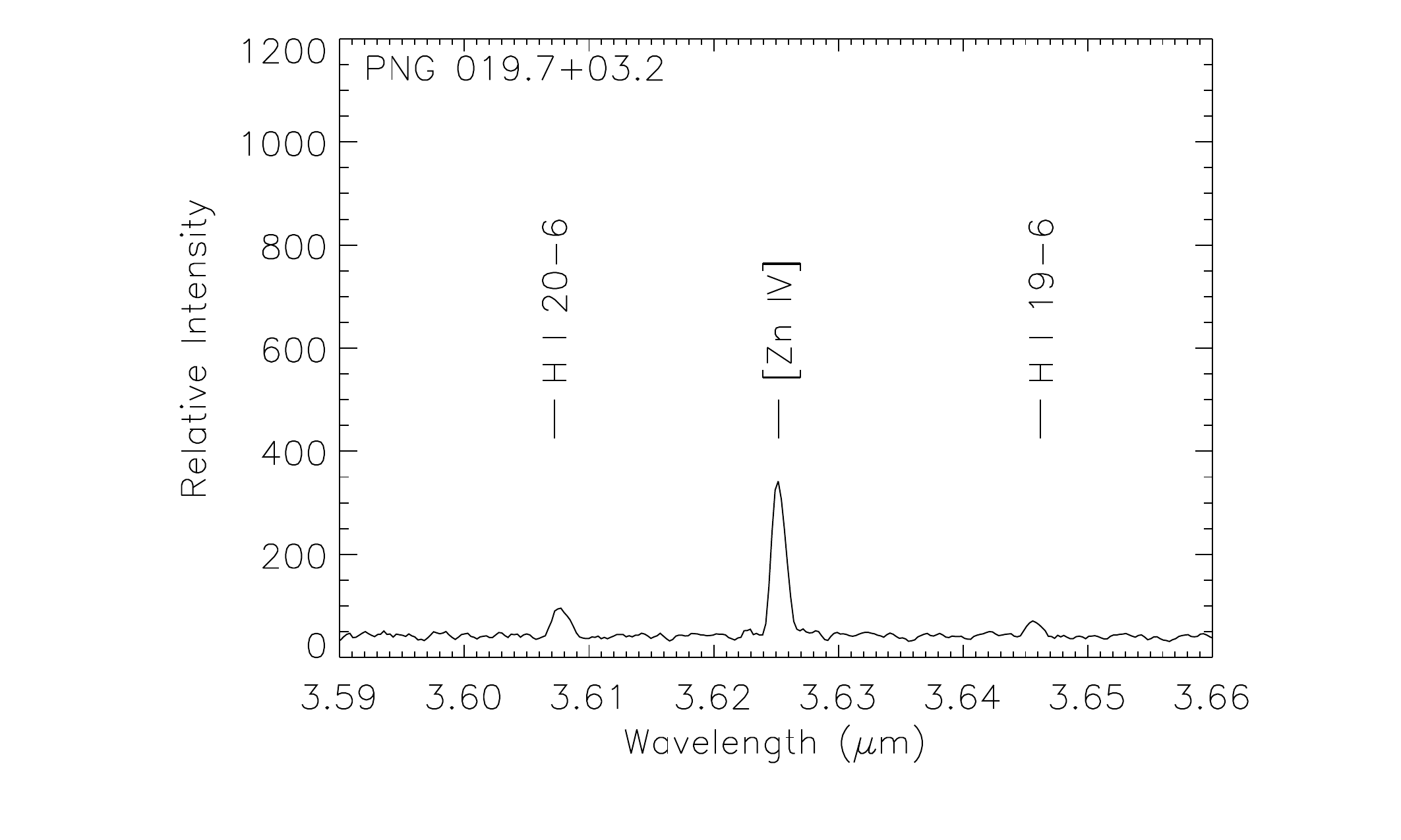}}
\subfigure{\includegraphics[trim=0cm 12cm 0cm 3cm, clip=true,scale=0.4]{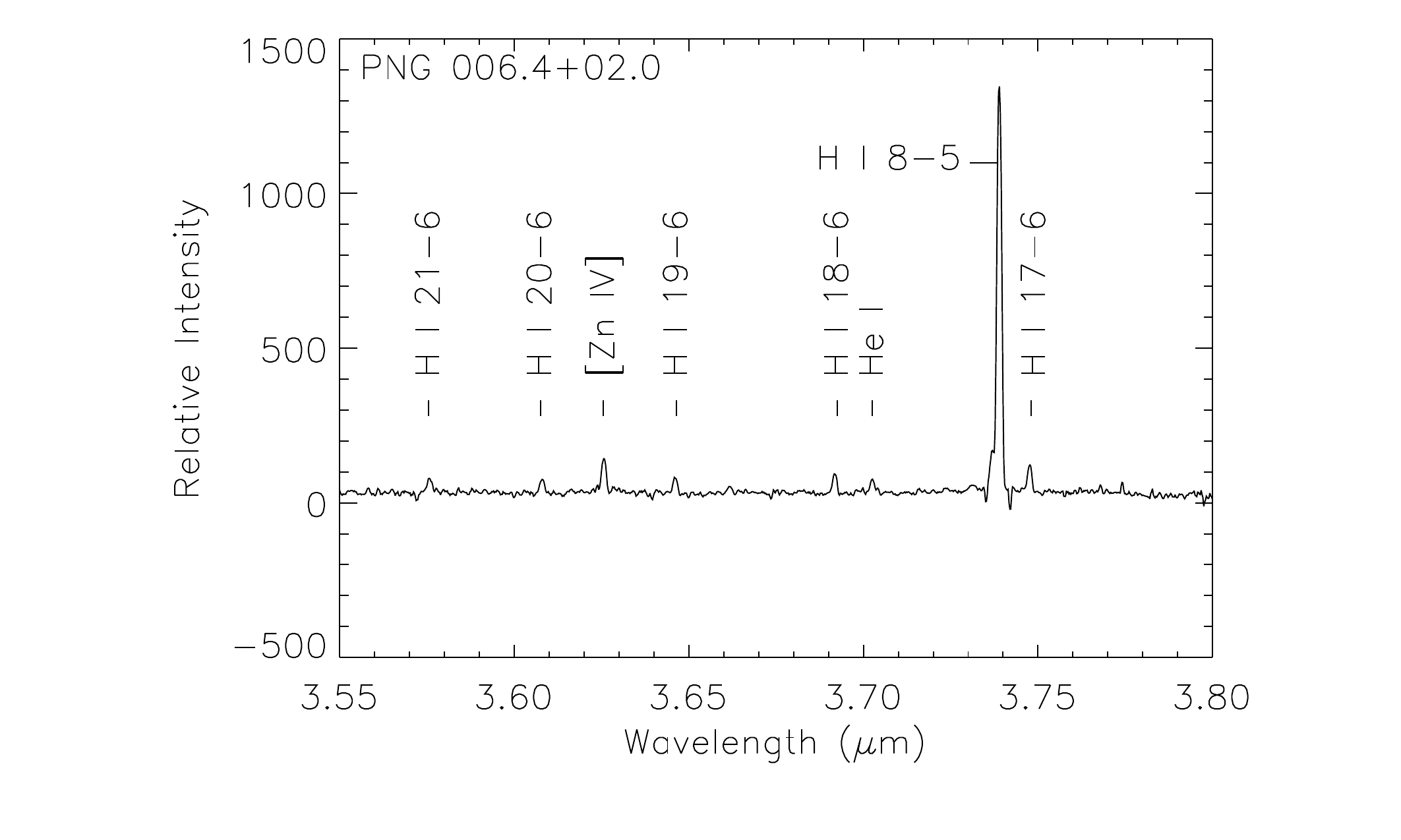}}
\subfigure{\includegraphics[trim=0cm 12cm 0cm 3cm, clip=true,scale=0.4]{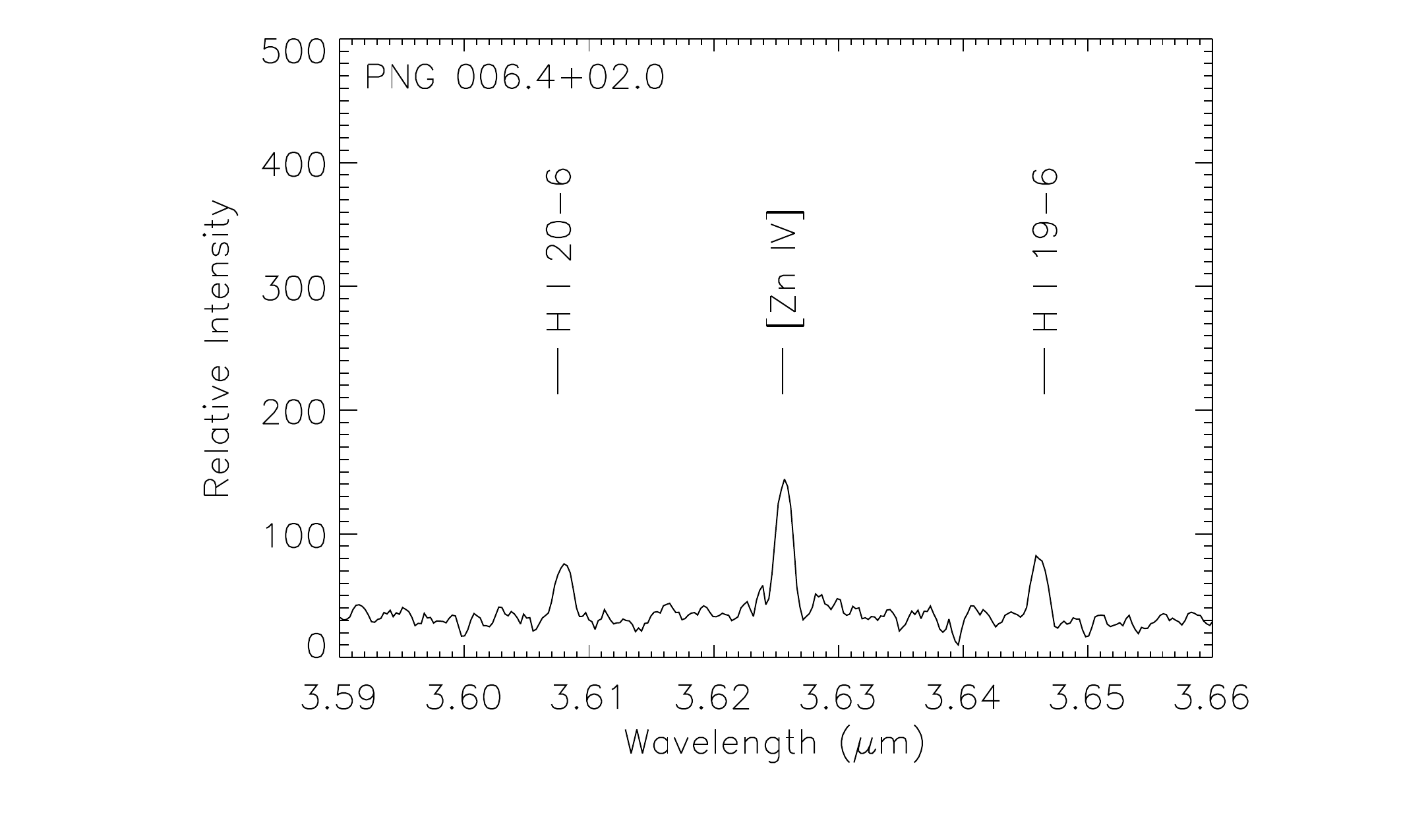}}
\subfigure{\includegraphics[trim=0cm 12cm 0cm 3cm, clip=true,scale=0.4]{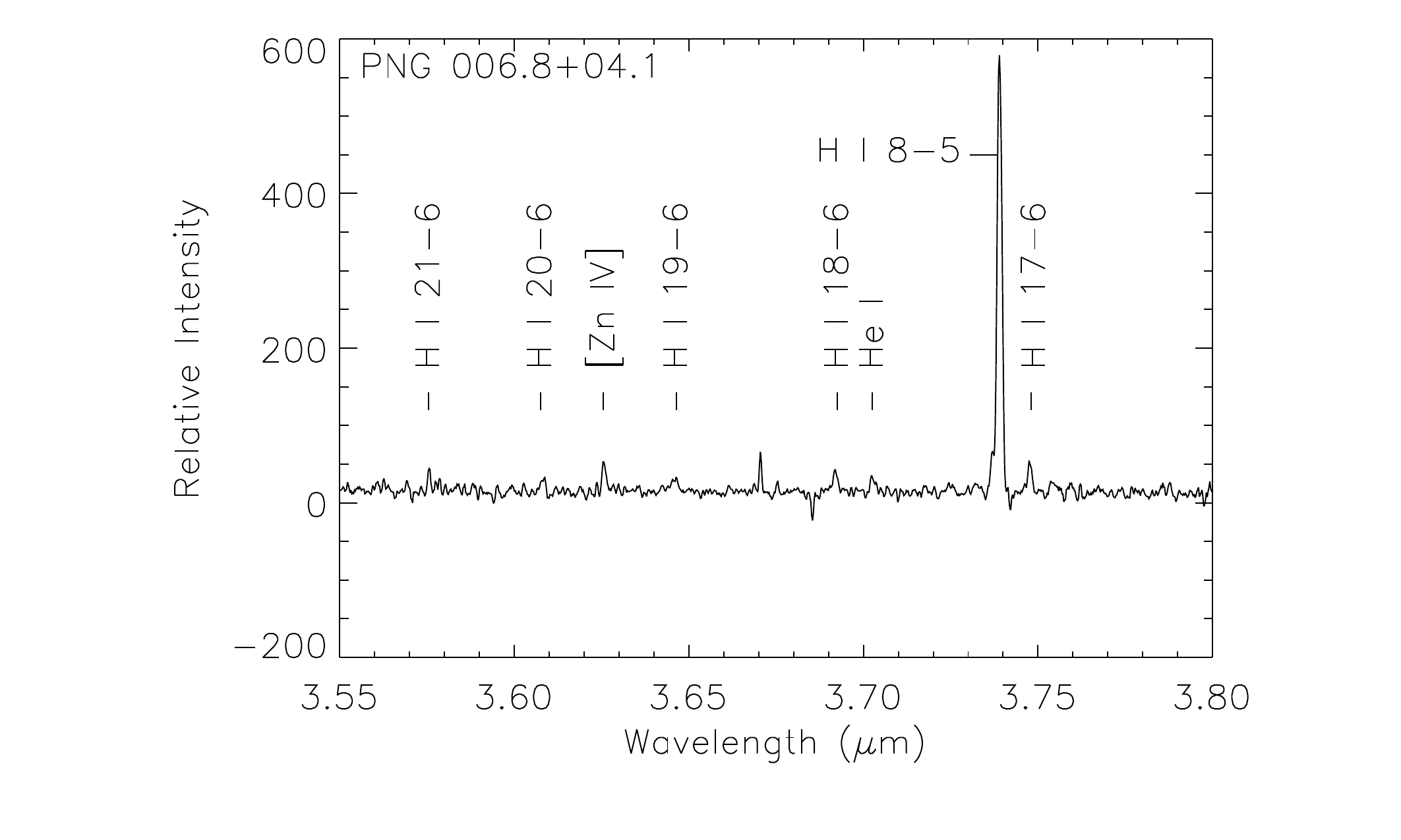}}
\subfigure{\includegraphics[trim=0cm 12cm 0cm 3cm, clip=true,scale=0.4]{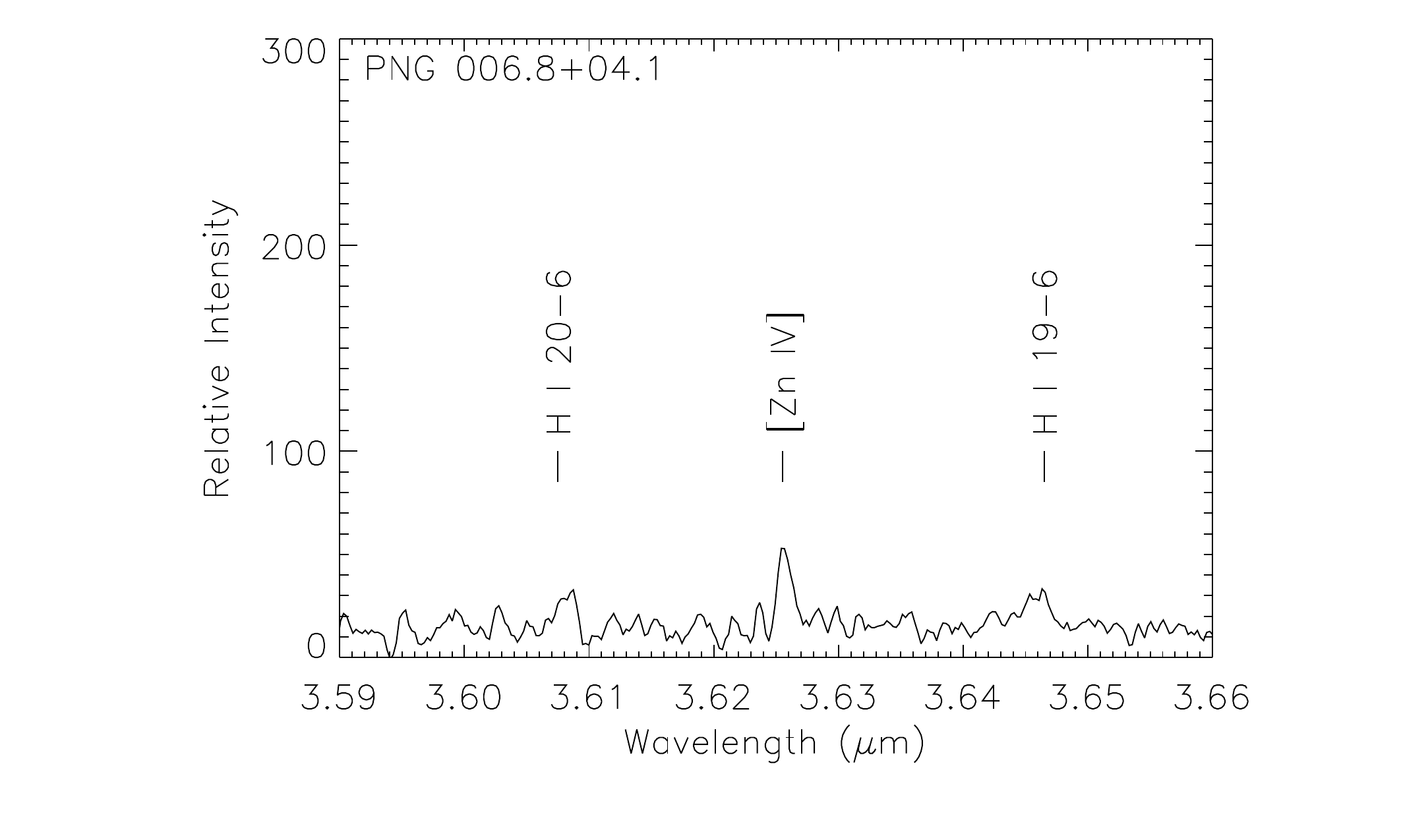}}
\subfigure{\includegraphics[trim=0cm 12cm 0cm 3cm, clip=true,scale=0.4]{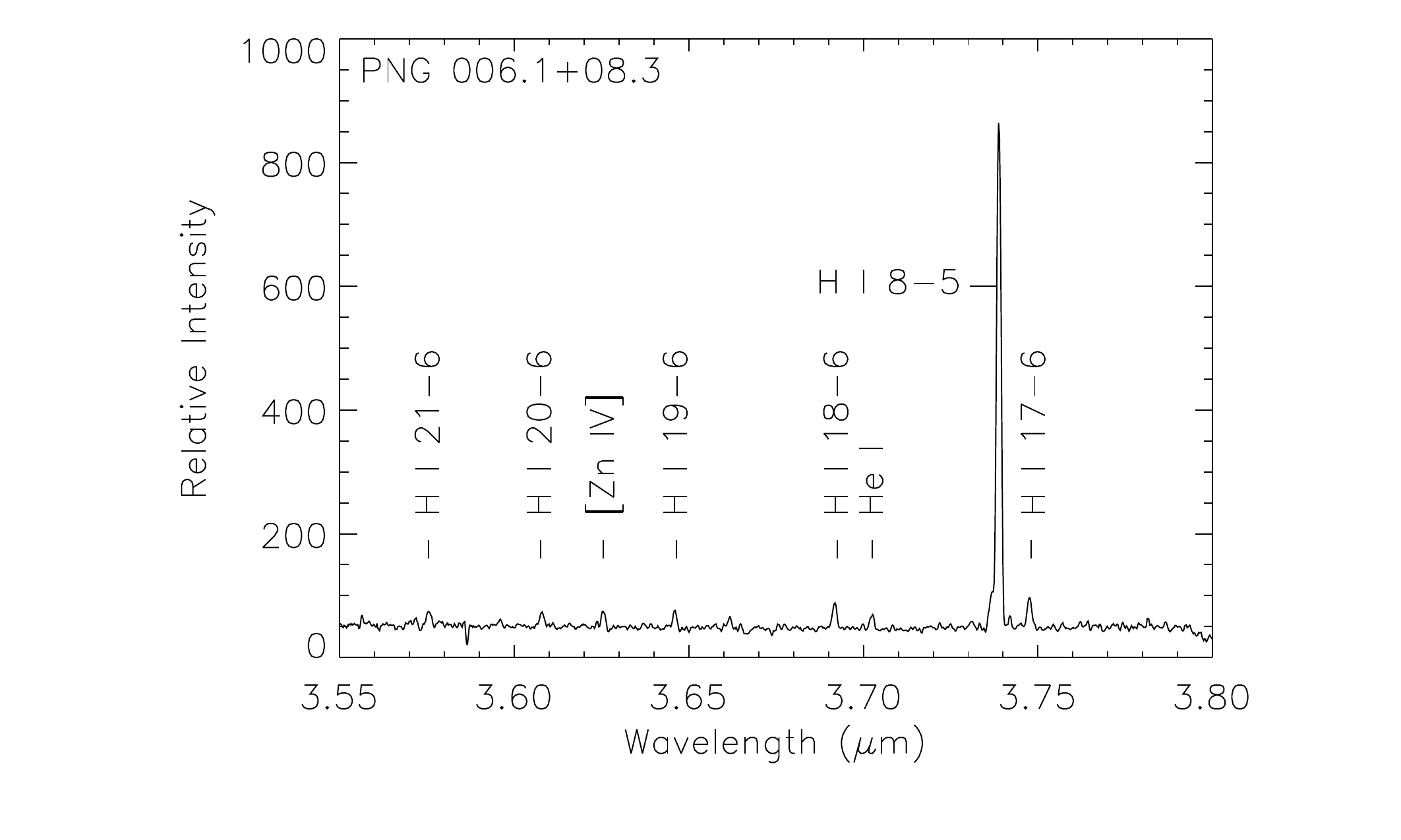}}
\subfigure{\includegraphics[trim=0cm 12cm 0cm 3cm, clip=true,scale=0.4]{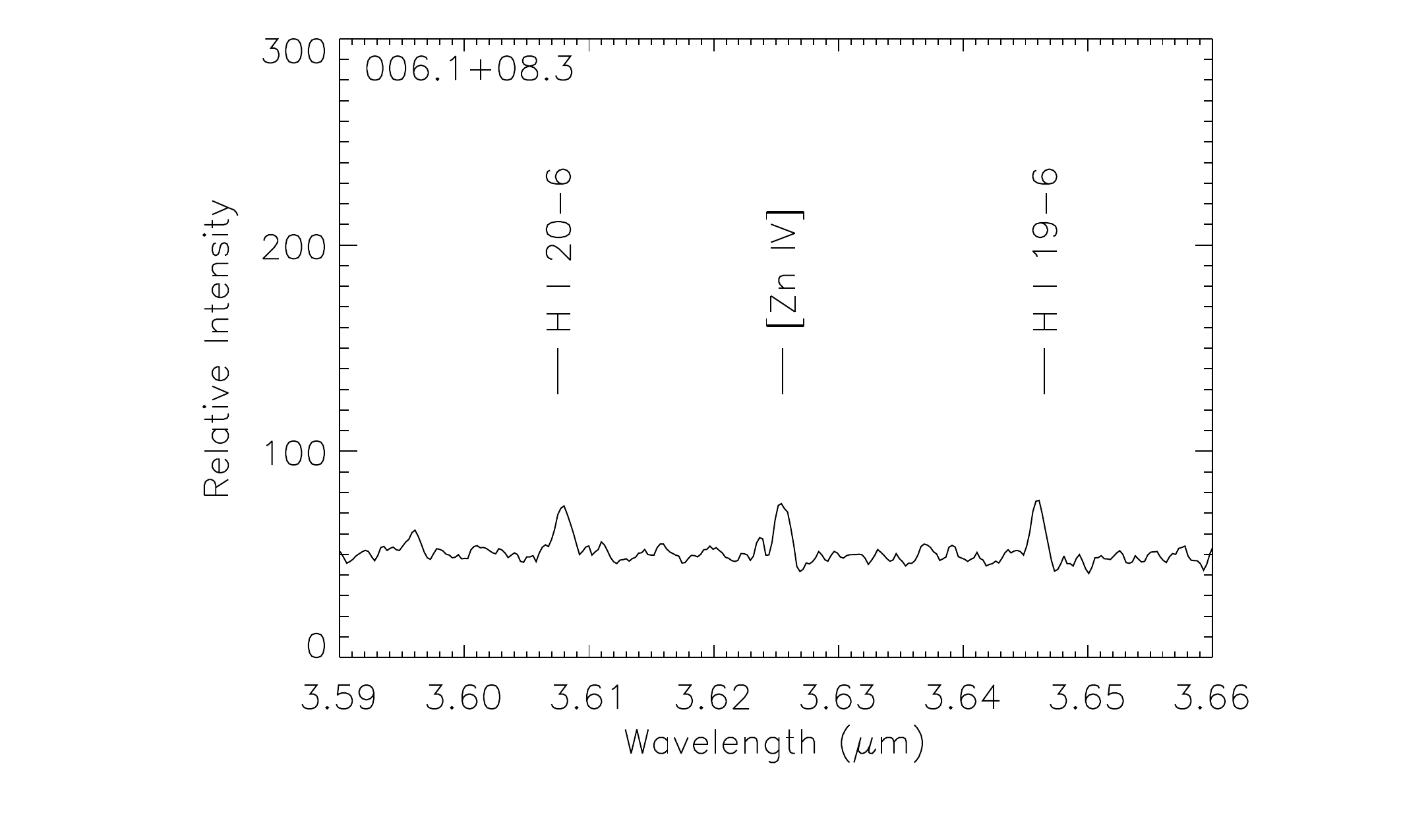}}
\caption{Spectra of all sources with detections or borderline detections of the  [Zn IV]  line.  The spectra have been background subtracted and subsequently smoothed, except that of PNG 355.4-02.4 which had a noise spike in close proximity to the [Zn IV] emission line and thus was not smoothed. }\label{spectra}
\end{figure*}
\begin{figure*}
\centering
\subfigure{\centering \includegraphics[trim=0cm 12cm 0cm 3cm, clip=true,scale=0.4]{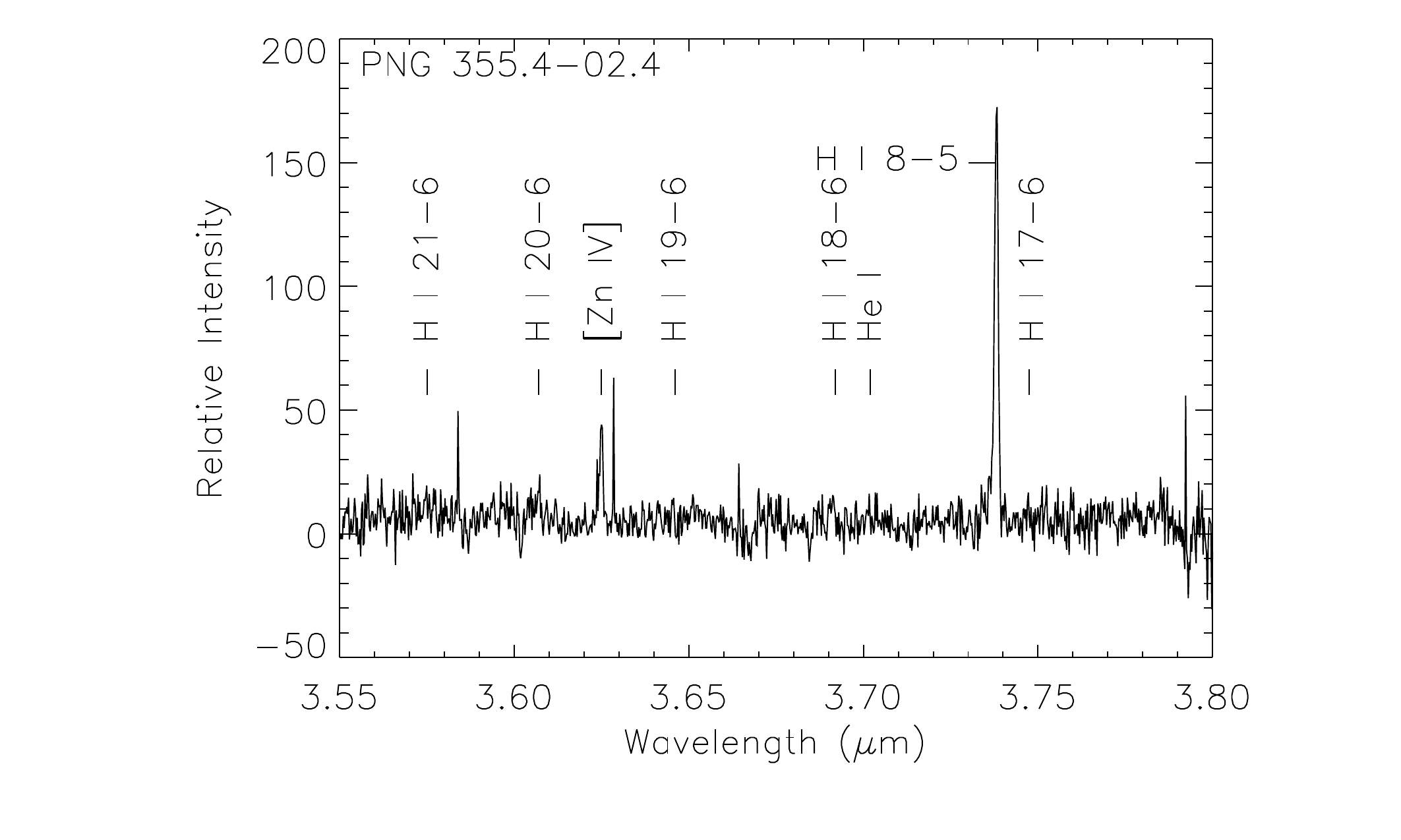}}
\subfigure{\centering \includegraphics[trim=0cm 12cm 0cm 3cm, clip=true,scale=0.4]{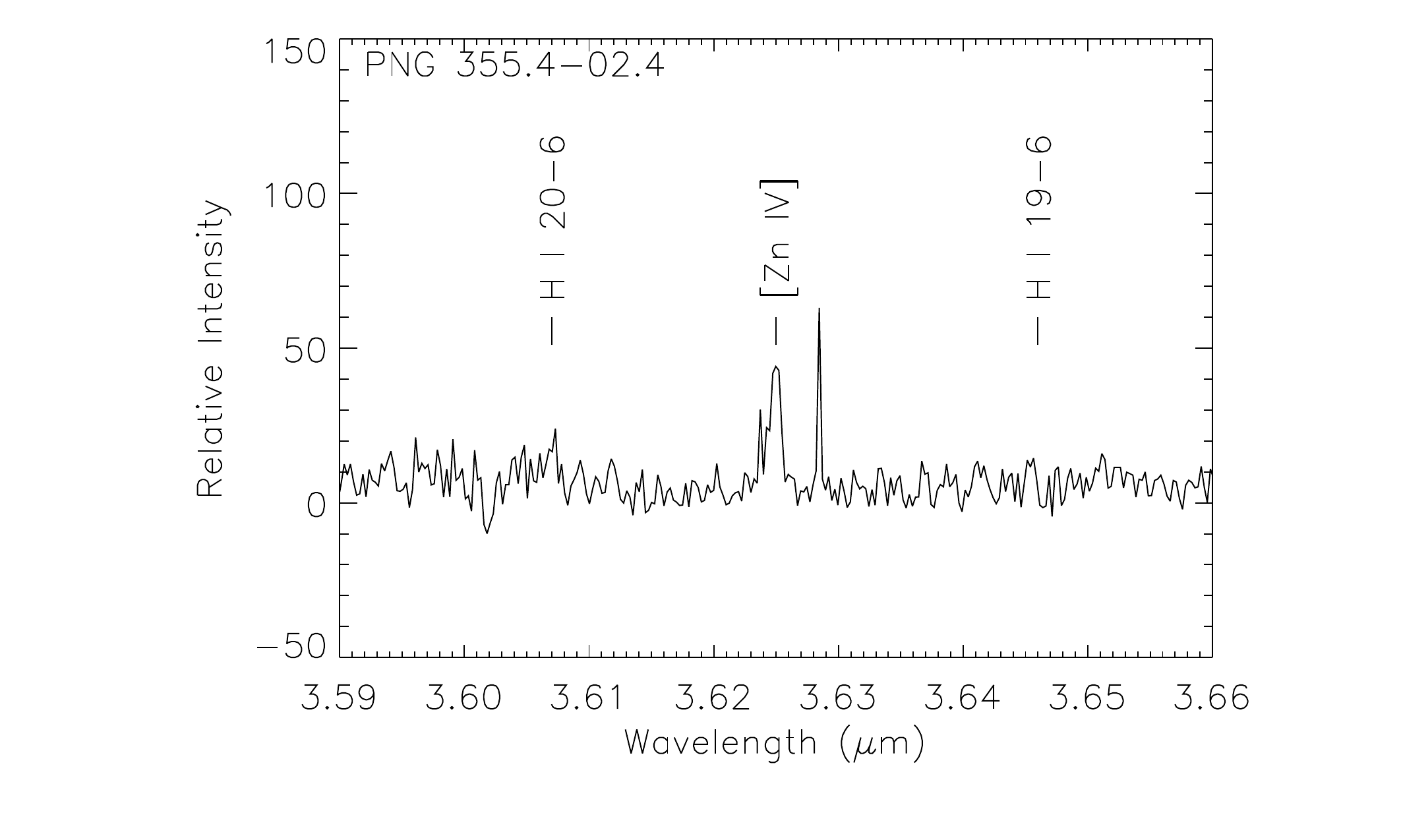}}
\subfigure{\includegraphics[trim=0cm 12cm 0cm 3cm, clip=true,scale=0.4]{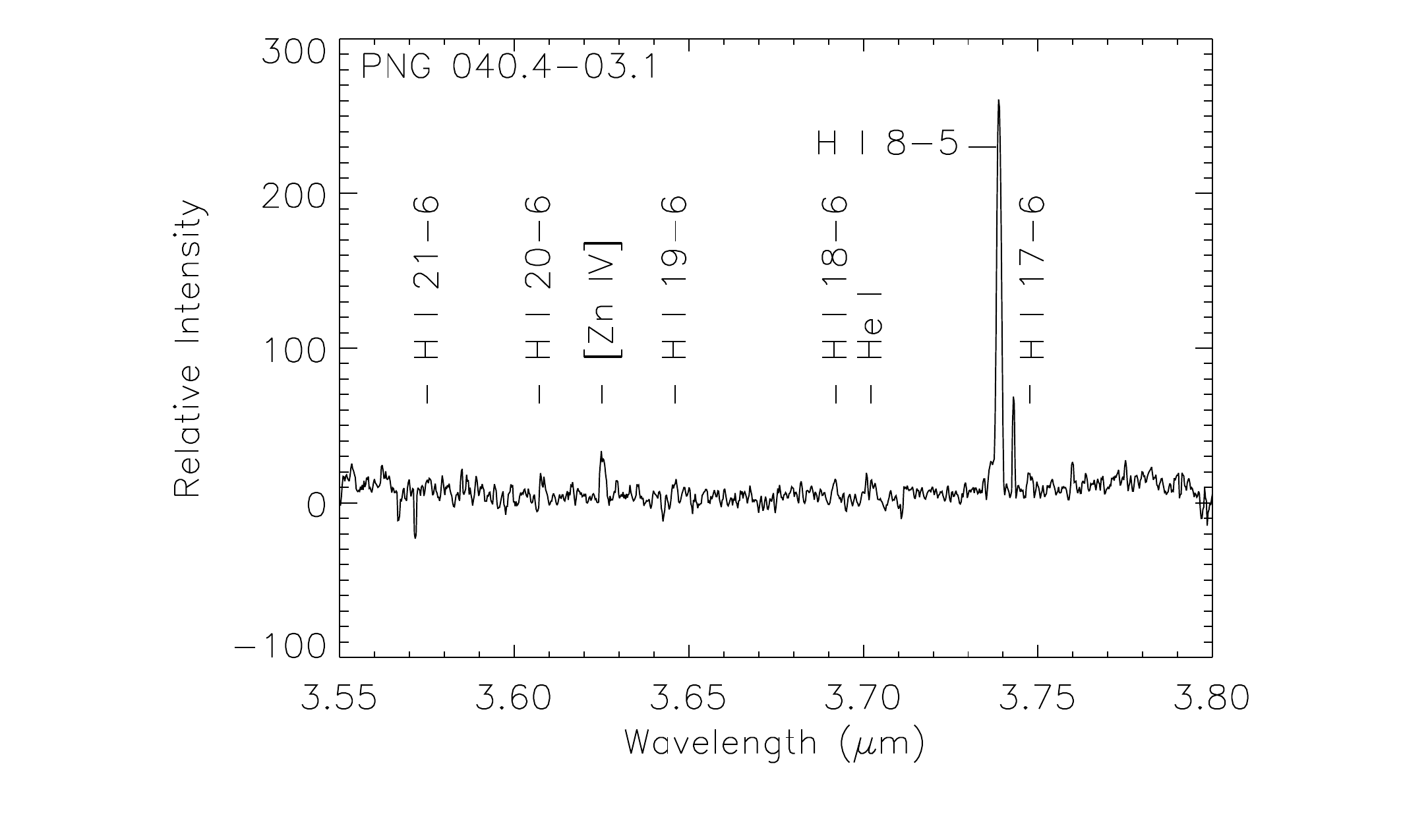}}
\subfigure{\includegraphics[trim=0cm 12cm 0cm 3cm, clip=true,scale=0.4]{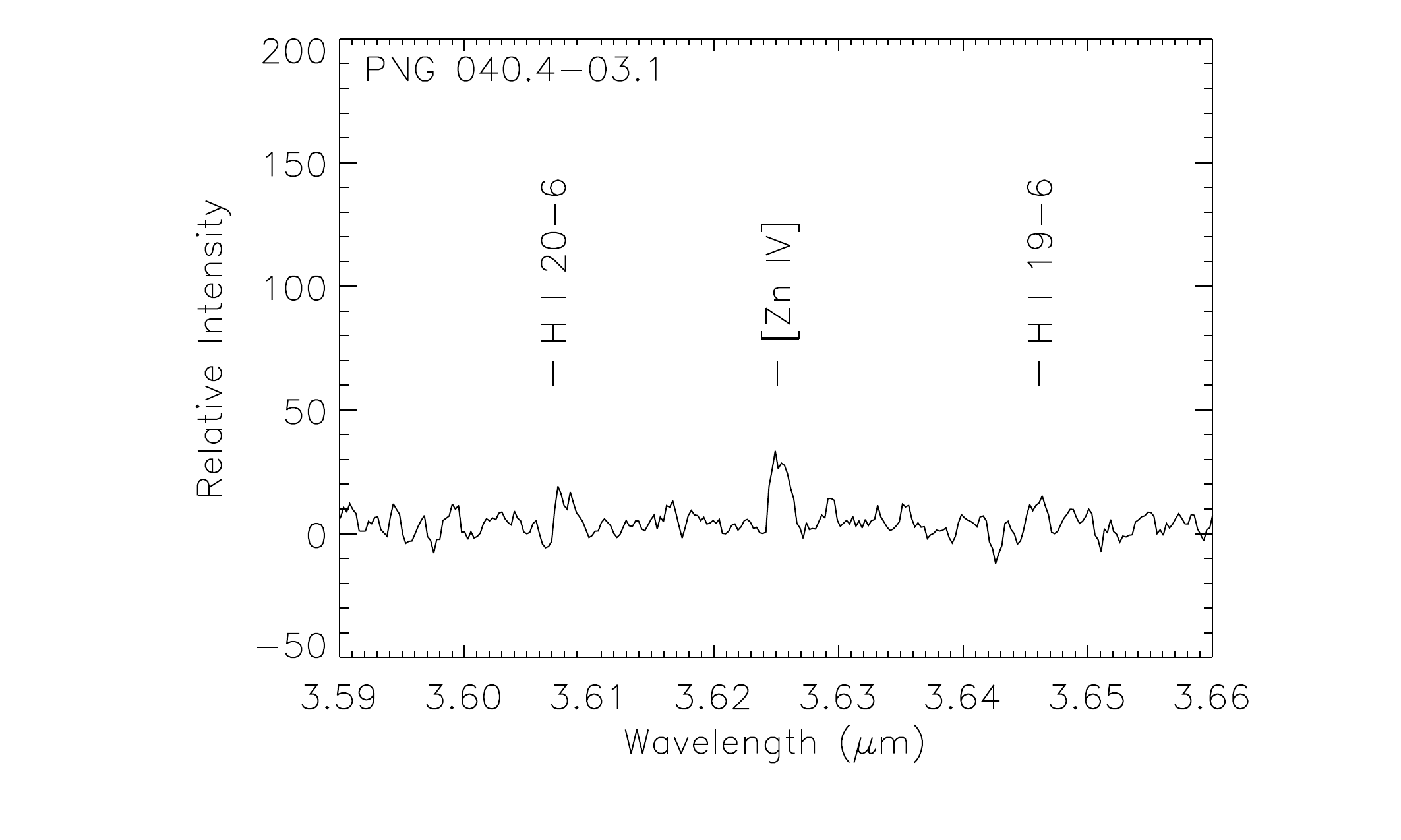}}

\subfigure{\includegraphics[trim=-0.5cm 12cm 0cm 3cm, clip=true,scale=0.4]{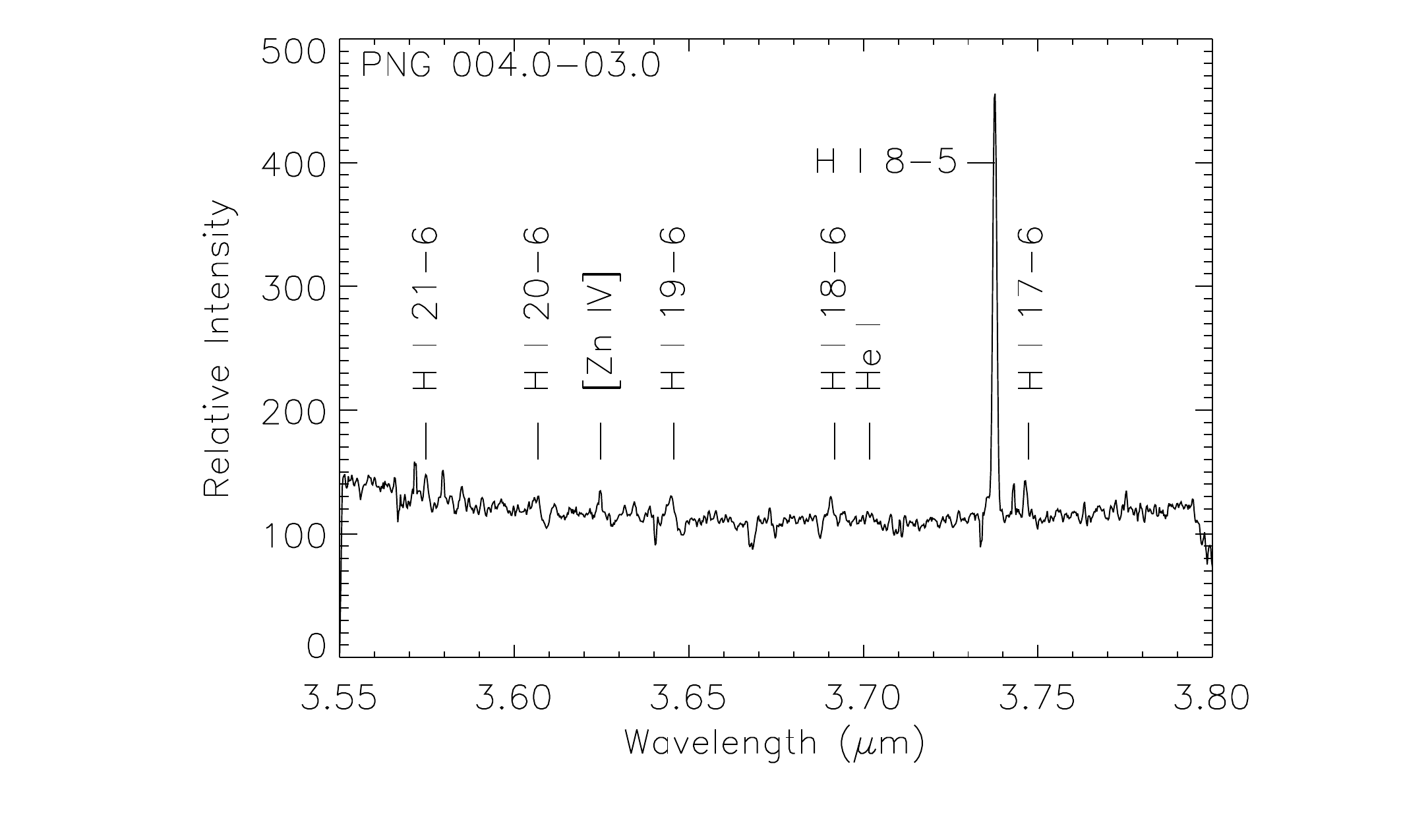}}
\subfigure{\includegraphics[trim=0cm 12cm 0cm 3cm, clip=true,scale=0.4]{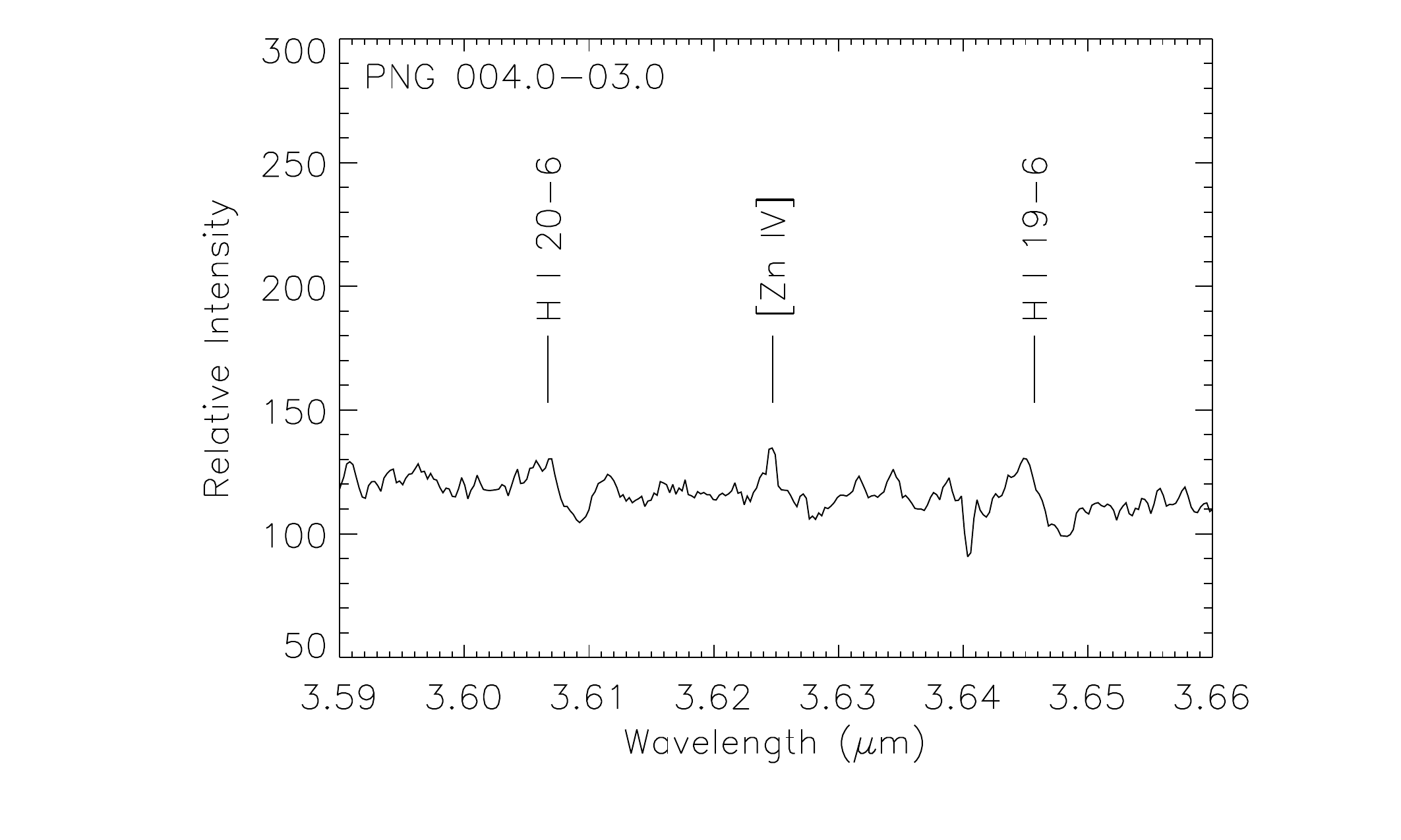}}
\contcaption{}
\end{figure*}

\subsection{Literature data}

The [Zn IV] 3.625 $\upmu$m emission line was first identified by \citet{Dinerstein2001} through observations of NGC 7027 and IC 5117. We include these observations, using the newly calculated values of the energy averaged collision strength, in order to compare the results from the new sample of nebulae to that of results obtained from well-studied nebulae, calculated using the same method as for the new sample.

The observations of \citet{Dinerstein2001} contained the 3.625 $\upmu$m emission line as well as two lines of hydrogen from the Humphreys series (n=20-6, 3.607 $\upmu$m and n=19-6, 3.646 $\upmu$m, hereafter H$_{20-6}$ and H$_{19-6}$), but did not include the stronger H$_{8-5}$ line which we use for the new sample due to its higher intensity.

\begin{figure}
\centering
\includegraphics[trim=0cm 13cm 0cm 2cm, clip=true,scale=0.4]{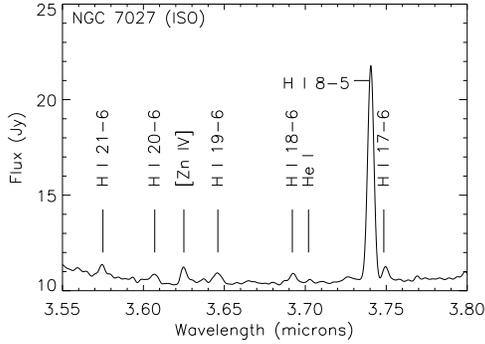}
\caption{\emph{ISO} spectrum of NGC 7027, TDT: 02401183, taken from \citet{Sloan2003}.}\label{ISOspectrum}
\end{figure}

The \emph{ISO} archive was also searched for further observations covering the 3.625 $\upmu$m wavelength range. NGC 7027 is the only planetary nebula observed at sufficient resolution to identify the [Zn IV] emission line: TDT=02401183 taken from \citet{Sloan2003}, shown in Fig. \ref{ISOspectrum}. Both the [Zn IV] and H$_{8-5}$ lines are visible, although at much lower resolution than the spectra in \citet{Dinerstein2001}. There were no new detections of [Zn IV] in other planetary nebulae found within the \emph{ISO} archive.

\section{Abundance derivation}\label{abd_der}

In order to calculate zinc abundances from observed line intensity ratios, a number of calculations must be carried out. First, the emissivities of the lines must be calculated, then the ionic abundances may be derived. The unobserved ions must be accounted for using a suitable ionisation correction factor before the elemental abundances can be calculated.

\subsection{Determination of physical parameters}

Basic information about each new source detected in zinc is shown in Table \ref{basic_info} as well as the two nebulae  previously observed by \citet{Dinerstein2001}. $T_\text{star}$ was quoted from the  second  reference in column 11 and the angular diameters are taken from \citet{Acker1992}. $T_\text{e}$(O III), $N_\text{e}$(S II), O/H and O$^{++}$/O were calculated from the optical spectra published in the  first  reference in column 11 using the {\sc{Nebular Empirical Analysis Tool}} ({\sc{NEAT}}, \citet{Wesson2012}).  {\sc{NEAT}} uses the \citet{Kingsburgh1994} ICF scheme. Further technical details of the code, including error propagation and sources of atomic data, may be found in \citet{Wesson2012}. The number of lines used as input varied from source to source, depending upon the available literature spectra. NGC 7027 had the largest input line list (227 lines in the wavelength range 3327.1--8727.2 $\AA$) and PNG $019.7+03.2$ had the smallest input line list (14 lines in the wavelength range 4340.4--7330.2 $\AA$). All lines published in the listed reference that are recognised by {\sc{NEAT}}  were used, and a full list of lines is available in each source's corresponding reference (column 11).

There was not a sufficiently complete set of emission line fluxes for PNG 019.7+03.2 in the literature to successfully apply the NEAT package, nor to derive $T_\text{e}$(O III) directly, so the value listed in Column  7   of Table \ref{basic_info} is $T_\text{e}$(N II). Using a published sample of 51 planetary nebulae from \citet{Gorny2009} with derived $T_\text{e}$(N II) and $T_\text{e}($O III), it is possible to extrapolate a $T_\text{e}($O III) from the available $T_\text{e}$(N II) value using a straight-line fit to their data. This would predict $T_\text{e}($O III)=1.05$\times10^4$ K for a $T_\text{e}$(N II)=1.09$\times10^4$ K. As these values are so similar, the effect on the emissivities is negligible, therefore $T_\text{e}$(N II) is used for the remainder of the paper. 

PNG 040.4 did not have sufficient optical data available in the literature to derive the required parameters. Therefore typical nebular parameters of 10$^4$ K and  10$^4$ cm$^{-3}$  were adopted for $T_\text{e}$(O III) and $N_\text{e}$(S II) respectively, and typical values of O/H and O$^{++}$/O for our disk nebula sample were assigned.

\subsection{Line emissivities}

Hydrogen emissivities are available as an extensive data list and an interpolation program from \citet{Storey1995}. The H$_{8-5}$ at 3.741 $\upmu$m is the strongest hydrogen line in our observed wavelength range, and is therefore the most appropriate line to use in the calculations. Using $T_\text{e}(\text{O III}$) and $N_\text{e}$(S II) listed in Table \ref{basic_info}, the interpolation code was used to give more accurate H$_{8-5}$ emissivities. For NGC 7027 and IC 5117, the H$_{20-6}$ and H$_{19-6}$ lines were used and their emissivities determined in the same manner as for H$_{8-5}$. 

[Zn IV] line emissivities are not available in the literature and must be calculated. Using the methodology described in \citet{Dinerstein2001} in the low-density limit and new collision strengths, $\Upsilon$, integrated over the electron energy distribution at temperature $T_\text{e}$ (Keith Butler, priv. comm.), the emissivity of the [Zn IV] line, $\varepsilon (\text{Zn IV})$, can be calculated using:

\begin{equation}
\varepsilon (\text{Zn IV}) = \text{exp}\left(-\frac{\Delta E_{ul}}{kT_e}\right)\frac{h\nu_{ul}\beta\Upsilon}{g_l\sqrt{T_e}}
\end{equation}

\noindent where $g_i$ is the statistical weight of level \emph{i} and $\Delta E_{ul}$ is the energy of the transition $u-l$. $\beta$ is defined as $\sqrt{(2\pi\hbar^4)/(m_{e}^3k)}=8.629\times10^{-6}$ (cgs units). 

The energy-averaged collision strengths were calculated over a grid of electron temperatures ranging between 5000 K and 25000 K (Keith Butler, priv. comm.). The energy-averaged collision strengths were interpolated to the electron temperatures of each source using a power law.

The use of the low density limit can be shown to be justified by deriving the critical density, $n_\text{crit}$, for which the rate of collisional de-excitations equals the rate of spontaneous decays:

\begin{equation}
\begin{split}
n_\text{crit}([\text{Zn IV}]) =\frac{A_{ul}}{q_{ul}} & =\left(\frac{A_{ul}g_u}{\beta\Upsilon_{ul}}\right)\sqrt{T_e} \\ & =1.4\times10^7\sqrt{\frac{T_e}{10^4\text{ K}}}\text{ cm}^{-3}
\end{split}
\end{equation}

\noindent where A$_\text{ul}$=0.43 s$^{-1}$, g$_u$=4 and $\Upsilon_\text{ul}$=1.4. As shown in Table \ref{basic_info}, the nebular densities are much lower than that of $n_\text{crit}$, thus the use of the low-density limit is well-justified.

\subsection{Ionic abundances}

\begin{table}
\caption{Telluric standards for each source, including their spectral type and effective temperatures, taken from the Hipparcos Catalogue.}\label{tellurics}
\begin{tabular}{c c c c}
PNG	&		Telluric Standard	& Spectral Type & T$_\text{eff}$ (K) \\
\hline\hline
004.0$-$03.0		&	Hip088012  &	 B3II/III & 17000 \\
006.1+08.3	&	Hip076243  &	 B6IV 	&	14000 \\
006.1+08.3	&	Hip085442  &	B9.5IV 	&	9790 \\
006.4+02.0	&	Hip085442 &	B9.5IV	&	9790 \\
006.8+04.1	&	HIP085355	& K3Iab	&	4420K \\
019.7+03.2	&	HIP085355	& K3Iab	&	4420K \\
040.4$-$03.1	&	Hip095732	& 	B6III &	17100 \\
355.4-02.4	&	Hip089439	& 	B0/1Ia/ab	& 	26000 \\
\end{tabular}
\end{table}

After measuring the integrated line intensity ratios of [Zn IV] and H$_{8-5}$, calculating the ionic abundance ratios is carried out using:

\begin{equation*}
\frac{\text{Zn}^{3+}}{\text{H}^+}=\frac{F(\text{[Zn IV]})}{F(\text{H}_{8-5})}\frac{\varepsilon(\text{H}_{8-5})}{\varepsilon(\text{Zn IV})}
\end{equation*}

\noindent where $F(i)$ is the flux of transition $i$, given in Table \ref{abs_abn}. Extinction effects are negligible in this wavelength range, so need not be considered.

\subsection{Ionisation correction factors}

To convert the ionic abundances into elemental abundances, it is necessary to correct for those ions of that species that are not seen. For hydrogen, it is assumed that all H is present as H$^+$, however this is not the case for Zn.

In order to examine the ionization correction factor required for zinc, as well as the dominance of the Zn$^{3+}$ ion, we have run a grid of {\sc{Cloudy}} (v 13.01) photo-ionization models \citep{Ferland2013}. The grid of models were given solar abundances and spherical geometry. Luminosity, initial radius and hydrogen density were set as  log($L_\text{tot}$[erg/s])=38, log($R_\text{initial}$[cm])=17  and log($n _\text{H}$[cm$^{-3}$]) = 4. The central star's radiation field was assumed to be a blackbody with temperatures ranging from $2\times 10^4$ K to $4\times 10^5$ K. 

\begin{figure}
\centering
\includegraphics[trim=2cm 13cm 2cm 2cm, clip=true,scale=0.45]{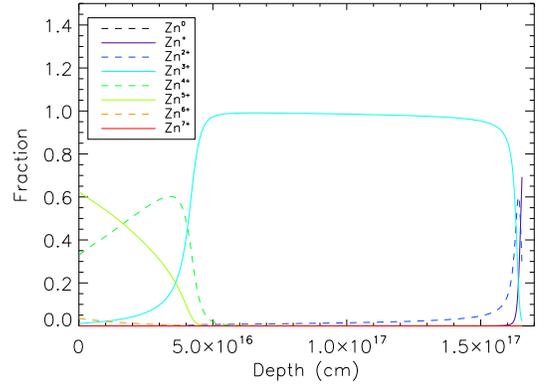}
\caption{Fractional ionisation abundance of zinc from {\sc{Cloudy}} models. This model: $T_\text{star}=10^5$K.}\label{cloudy_zn}
\end{figure}

At central star temperatures between $6\times 10^4$ K and $1\times 10^5$ K, Zn$^{3+}$/Zn $> 0.9$ at most depths, where depth is defined as the distance from the illuminated face of the nebular cloud. When the depth into the cloud is small, the fractions of zinc in other ions becomes significant, as shown in Fig. \ref{cloudy_zn}. The dominance of Zn$^{3+}$ also does not hold when examined at temperatures smaller than $6\times 10^4$ K and greater than $1\times 10^5$ K. 

\begin{figure}
\centering
\includegraphics[trim=2cm 13cm 2cm 2cm, clip=true,scale=0.45]{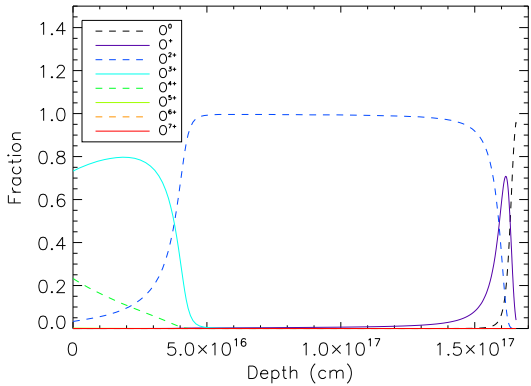}
\caption{Fractional ionisation abundance of oxygen from {\sc{Cloudy}} models. This model: $T_\text{star}=10^5$K.}\label{cloudy_oxy}
\end{figure}

\begin{figure}
\centering
\includegraphics[trim=2cm 13cm 2cm 2cm, clip=true,scale=0.44]{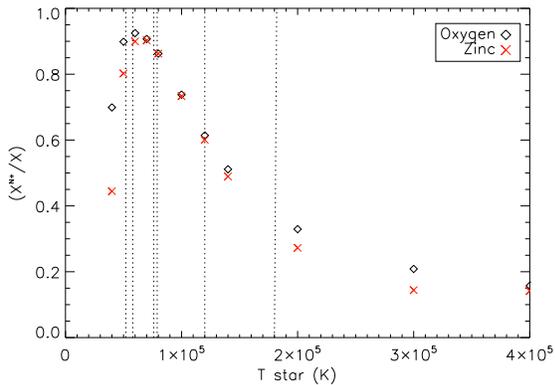}
\caption{ Fractional abundances of O$^{++}$ and Zn$^{3+}$ over the central star temperature range $4\times 10^4$ K to $4\times 10^5$ K. Dotted vertical lines indicate the central star temperatures of the sample nebulae.}\label{cloudy_allt}
\end{figure}

These models also show that over the range of central star temperatures $5\times 10^4$ K to $1.5\times 10^5$ K, the fraction of zinc in the Zn$^{3+}$ ion is closely mapped by the fraction of oxygen present as O$^{++}$, which is the dominant oxygen ion over the same temperature range, as shown in Fig. \ref{cloudy_oxy} and \ref{cloudy_allt}. O$^{++}$/O has been determined for the majority of our sample from optical observations and all bar one of the sample of planetary nebulae central stars lie within this temperature range. Therefore, the ionisation correction factor for Zn$^{3+}$ can be taken as ($\text{O}^{++}/\text{O})^{-1}$ and the values are listed in Table \ref{basic_info} for each nebula.

\begin{figure}
\centering
\includegraphics[trim=2cm 13cm 2cm 2cm, clip=true,scale=0.45]{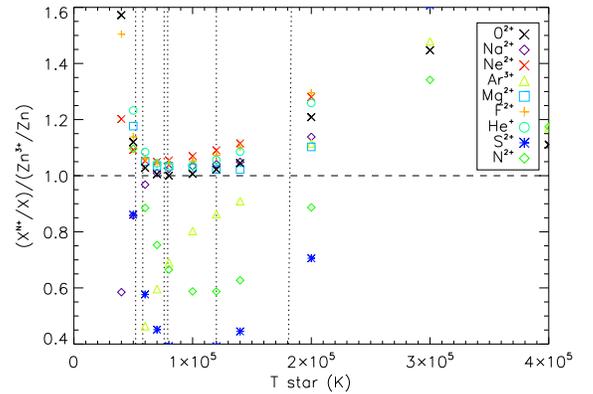}
\caption{ Fractional abundances of a variety of species in comparison to the fractional abundance of Zn$^{3+}$ over the central star temperature range $4\times 10^4$ K to $4\times 10^5$ K. The species included are those that most closely follow the fractional abundance of Zn$^{3+}$, except Fe ions. The dotted vertical lines indicate the central star temperatures of the sample nebulae.}\label{all_elements}
\end{figure}

\citet{Dinerstein2001} use Ar/Ar$^{3+}$ as the ionisation correction factor rather than O/O$^{++}$. Our {\sc{Cloudy}} models show that O$^{++}$/O follows Zn$^{3+}$/Zn  more closely than Ar$^{3+}$/Ar does, hence our choice of species.  Additionally, over the temperature range $6\times10^4$ K to $1.2\times10^5$ K, the models show that Zn$^{3+}$/Zn is most closely mapped by O$^{++}$/O in comparison to all other species included in the model nebula.  This is shown in Fig. \ref{all_elements}.

\subsection{Elemental abundances}

From the calculated ionic abundances and using O$^{++}$/O to correct for the unobserved ions of zinc, the Zn elemental abundance is simply:
\begin{equation}
\frac{\text{Zn}}{\text{H}}=\frac{\text{Zn}^{3+}}{\text{H}^+}\frac{\text{O}}{\text{O}^{++}}
\end{equation}

\noindent Outside of the temperature range $5\times 10^4$ K to $1.5\times 10^5$ K, applying this correction will lead to inaccurate results where the abundance of Zn will be significantly underestimated. Zn/H may be converted to Zn/O using the O/H values shown in Table \ref{basic_info}.  As discussed earlier, the value of Zn/Fe has essentially the solar value for stars with $[{\rm{Fe/H}}] > -2$ \citep{Saito2009, Umeda2002, Sneden1991}, which includes all of our targets.   Thus, it is possible to convert from [Zn/H] and [Zn/O] to [Fe/H] and [Fe/O] using the solar ratio of Zn/Fe.

\section{Results}

The results are shown in Table \ref{abs_abn} in both logarithmic and linear forms. The solar elemental abundances are taken from \citet{Asplund2009}. The results from the {\emph{ISO}} observations are not displayed in Table \ref{abs_abn} as they are within errors of those determined from the observations in \citet{Dinerstein2001}, but due to their lower resolution had significantly larger uncertainties.

\begin{table*}
\centering
\caption{Abundances of zinc and iron with respect to hydrogen and oxygen for the new sample and the literature sample. Iron abundances may be calculated from the zinc abundances using the solar ratio of Zn/Fe.  Flux ratios are listed in column 3 and are given as the flux of the [Zn IV] line with respect to to the flux of the H line used, H$_{8-5}$ unless otherwise indicated.  Emissivities are quoted in erg s$^{-1}$ cm$^{-3}$.}  \label{abs_abn}
\begin{tabular}{c c c c c c c c}
 & & & & \\
\multirow{2}{*}{PNG} & $\varepsilon$([Zn IV])	&	F$_{[\text{Zn IV}]}$/F$_\text{H}$ & $\text{Zn}^{3+}/\text{H}^+$ & Zn/H & Zn/O & \multirow{2}{*}{[Zn/H]} & \multirow{2}{*}{[O/Zn]} \\
	& ($\times10^{-21}$)	& ($\times10^{-2}$)	& ($\times10^{-8}$)		& ($\times10^{-8}$)	& ($\times10^{-5}$)	& 	& 	\\
\hline\hline
$004.0-03.0$ & $7.6\pm{1.8}$ & $4.4\pm{1.5}$ & $0.35^{+0.16}_{-0.15}$ & $0.39^{+0.28}_{-0.38}$ & $13^{+10}_{-13}$ & $-1.0^{+0.2}_{-0.3}$ & $-0.2^{+0.2}_{-0.3}$ \\
$006.1+08.3$ & $7.1\pm{1.4}$ & $2.6\pm{0.4}$ & $0.5\pm{0.1}$ & $0.5\pm{0.1}$ & $1.5\pm{0.4}$ & $-0.9\pm{0.1}$ & $0.70\pm{0.1}$ \\
$006.4+02.0$ & $6.6\pm{1.3}$ & $8.1\pm{1.1}$ & $2.2\pm{0.6}$ & $2.3\pm{0.9}$ & $3\pm{1}$ & $-0.2\pm{0.1}$ & $0.4\pm{0.2}$ \\
$006.8+04.1$ & $6.8\pm{1.3}$ & $5.6\pm{0.8}$ & $1.3\pm{0.3}$ & $1.3\pm{0.3}$ & $2.2\pm{0.7}$ & $-0.4\pm{0.1}$ & $0.5\pm{0.1}$ \\
$019.7+03.2$ & $7.2\pm{1.4}$ & $17.4\pm{2.5}$ & $2.8\pm{0.7}$ & $3.5\pm{0.9}$ & $9\pm{3}$ & $-0.02\pm{0.10}$ & $-0.08\pm{0.10}$ \\
$040.4-03.1^{a}$ & $7.2\pm{1.6}$ & $10.0\pm{1.4}$ & $1.8\pm{0.8}$ & $2\pm{1}$ & $6\pm{5}$ & $-0.2\pm{0.2}$	& $0.1\pm{0.3}$ \\
$355.4-02.4$ & $6.9\pm{1.4}$ & $24.4\pm{3.5}$ & $5.5\pm{1.4}$ & $7.5\pm{1.9}$ & $12\pm{3}$ & $0.3\pm{0.1}$ & $-0.2\pm{0.1}$ \\
\hline
NGC 7027 & $7.4\pm{1.5}$ & $149\pm{7}^\text{b}$ & $0.88\pm{0.18}$  & $1.3\pm{0.3}$ & $3.4\pm{0.8}$ & $-0.44\pm{0.08}$	& $0.34\pm{0.09}$ \\
IC 5117 & $7.4\pm{1.5}$ & $260\pm{50}^\text{b}$ & $1.7\pm{0.4}$ & $1.8\pm{0.5}$ & $7\pm{2}$ & $-0.31\pm{0.10}$	& $0.06\pm{0.12}$ \\
\end{tabular}
\bigskip
{\begin{flushleft}
a: values are an estimate using average values for oxygen ratios in calculations where literature values were unavailable

b: flux are given for the H$_{19-6}$ transitions and abundances are the mean of those derived from the H$_{19-6}$ and H$_{20-6}$ transitions, taken from \citet{Dinerstein2001}.
\end{flushleft}}
\end{table*}

\subsection{Uncertainty considerations}

All uncertainties associated with the physical conditions and chemical composition of the sample nebulae are shown in Table \ref{basic_info}. The uncertainties on the measured flux ratio are given in Table \ref{abs_abn}. These errors have all been combined according to standard error propagation to give the uncertainties on the final abundances.

For PNG 040.4--03.1, we have adopted values of $\pm{0.2}$ and $\pm{0.3}$ dex for [Zn/H] and [O/Zn] ratios respectively.

There are also further uncertainties associated with the use of ionisation correction factors. The ionisation correction factor O/O$^{++}$ agrees to within 2\% with the value of Zn/Zn+3 predicted by the photoionization models, except for NGC 7027 and PNG 019.7+3.2, for which the uncertainty introduced by the ionisation correction is about 10\%, due to the very high and low central star temperatures respectively.  The calculated collision strength for [Zn IV] is estimated to be accurate to 20$\%$ (Keith Butler, priv. comm.), and the line emissivity should have a similar uncertainty. The uncertainties in the emissivities of the optical oxygen lines are likely to be greater due to their strong temperature dependences.

The sensitivity of the results on the derived physical input parameters can be examined by comparing the results derived using the parameters calculated by the {\sc{Neat}} code with those calculated from literature parameters. In general, the results of [Zn/H] agree to less than 0.08 dex, which is within the calculated uncertainties of the results. The only nebula with a difference in [Zn/H] greater than 0.08 dex is NGC 7027, with a difference of 0.14 dex. The results of [O/Zn] agree to within 0.09 dex for all nebulae except PNG 006.4+02.0 which has a discrepancy of 0.3 dex. This difference is primarily caused by an increase in the value of $\rm{O}^{++}/\rm{O}$ as derived through {\sc{Neat}} in comparison to that value derived in the literature. In general, the results are consistent to within calculated uncertainties.

The H$_{8-5}$ line is asymmetrical in profile and potentially caused by a blend with a significantly weaker He I emission line. In order to assess the effect this has on the calculated abundances, we have calculated abundance ratios for the four highest signal-to-noise spectra (namely PNG $006.1+08.3$, PNG $006.4+02.0$, PNG $006.8+04.1$ and PNG $019.7+03.2$) from both the peak intensities of the H$_{8-5}$ and [Zn IV] line and also using the integrated intensities of the H$_{18-6}$ lines. The results are not significantly different from those derived using the integrated intensities of the H$_{8-5}$ line. In general, [Zn/H] is reduced by 0.1 dex, but the abundance ratios of some nebulae are entirely unaffected (e.g. PNG $019.7+03.2$). The abundances derived for NGC 7027 and IC 5117 are unaffected due to the H$_{19-6}$ and H$_{20-6}$ lines being used in their calculations.

\section{Discussion}

All calculated ionic and elemental abundance ratios are shown in Table \ref{abs_abn}. Our results show a general trend of sub-solar Zn/H, with the lowest being 10\% that of solar.

We have derived [Zn/H] and [O/Zn] in order to calculate the metallicities of the precursor stars of the planetary nebulae in our sample and to determine whether $\alpha$ elements such as oxygen are enriched relative to Solar. A range of O/Zn is shown, with four of the sample nebulae showing enhancement of O/Zn in comparison to solar values.

PNG 004.0--03.0 shows sub-solar [O/Zn], however the uncertainties associated with this nebula are large, putting the solar value well within one sigma. [Zn/H] for this nebula is the most sub-solar at -1.0, five sigma away from Solar.   Additionally, as discussed in \citet{Gesicki2010}, this nebula contains a high density inner region which is likely an opaque disk. This may be the cause of the unusually low calculated values.

PNG 355.4--02.4 showed a very intense  [Zn IV]  emission line and has the highest F([Zn IV])/F(H$_{8-5}$) of all the sample, leading to the highest [Zn/H] in the sample,   suggesting that this nebula may simply be more metal-rich than the Sun . In addition to the high F([Zn IV])/F(H$_{8-5}$) ratio, no Humphreys emission lines were detected in PNG 355.4--02.4. It is possible that this nebula could be hydrogen-poor rather than zinc-rich,  although the low signal-to-noise ratio in this spectrum may be the cause of the non-detections of the Humphreys emission lines . The 3.625 $\upmu$m line has several noise spikes in close proximity, and it is possible a noise spike could be overlaid on the emission line, significantly increasing the measured zinc abundance which would also account for the high value of [Zn/H].

PNG 006.1+08.3, PNG 006.4+02.0 and PNG 006.8+04.1 show $\alpha$ enhancement, with oxygen being enhanced over zinc by factors of 5, 2.5 and 3 respectively. PNG 006.4+02.0 and PNG 006.8+04.1 both have significantly sub-solar [Zn/H]. PNG 006.1+08.3, on the other hand, shows a [Zn/H] ratio that is more in line with Solar. These results imply that these three nebulae are, to varying degrees, enhanced in $\alpha$-elements over zinc. 

PNG 019.7+03.2, a disk nebula, is consistent with Solar for both Zn/H and O/Zn. 

Due to the lack of optical spectroscopy for the probable disk planetary nebula PNG 040.4-0.3, it was necessary to assume values for its electron temperature, electron density, and ionic and elemental oxygen abundances. Its Zn/H and O/Zn values are consistent with Solar, within estimated errors, however the results should be recalculated if optical spectra for this nebula become available.

The well-studied disk nebula NGC 7027 displays Zn/H at approximately 1/3 that of the Solar value, whilst being enhanced in O/Zn by a factor of 2. This deviation from Solar may be explained in part by its large size of 14 arcseconds. The zinc observations and optical observations were taken with different slit widths at different orientations, and therefore will not trace the same regions of gas. In such a large nebula, this effect could be significant and could cause the underestimation of the zinc abundance by the slit losses being different for different elements. For the majority of the other nebulae, the slit and nebulae have similar sizes meaning that this is less likely to cause discrepancies.

IC 5117 has a lower Zn/H ratio than solar by a factor of two, whilst being consistent with solar for O/Zn.

Two of the sample nebulae, PNG $006.1+08.3$ and PNG $006.8+04.1$, show significant enhancements in [O/Zn] such that when errors are taken into account, they still lie at values greater than 0.2 over solar. Both nebulae have central star temperatures of $7.9\times10^4$ K, putting them within the temperature range where $\rm{O}^{++}/\rm{O}$ maps $\rm{Zn}^{3+}/\rm{Zn}$ in model nebulae to within $\sim2\%$.

\subsection{Metallicity as a function of galactocentric distance}

The planetary nebulae in this sample are at a variety of Heliocentric distances and Galactic latitude and longitudes. Transforming these parameters into a Galactocentric distance can be simply executed via geometry. The distance between the Sun and the Galactic centre has been taken to be 8.5kpc.

The l and b coordinates for all sources have been taken from \citet{Cutri2003} and the Heliocentric distances from \citet{Stanghellini2010}, and are shown, along with their calculated Galactocentric distances (R$_{\rm{G}}$), in Table \ref{GC_table}.

\begin{table}
\centering
\caption{Heliocentric (R) and calculated Galactocentric distances (R$_{\rm{G}}$) and l and b coordinates for all sources. Heliocentric distances and calculated Galactocentric distances are accurate to $20\%$.}\label{GC_table}
\begin{tabular}{c c c c c}
PNG & R (kpc) & l (deg) & b (deg) & R$_{\rm{G}}$ (kpc) \\
\hline
$004.0-03.0$ & 8.4 & 4.087 & -3.005 & 0.79  \\
$006.1+08.3$ & 8.0 & 6.188 &  +8.362 & 1.6  \\
$006.4+02.0$ & 5.2 & 6.455 & +2.015 & 3.4  \\
$006.8+04.1$ & 6.8 & 6.805 & +4.160 & 2.0  \\
$019.7+03.2$ & 5.8 & 19.752 & +3.273 & 3.6  \\
$040.4-03.1$ & 9.5 & 40.441 & -3.157 & 6.0  \\
$355.4-02.4$ & 5.8 & 355.444 & -2.467 & 2.8  \\
NGC 7027 & 1.1 & 84.930 & -3.496 & 8.5 \\
IC 5117 & 8.2 & 89.873 & -5.134 & 12  \\
\end{tabular}
\end{table}

\begin{figure}
\includegraphics[trim=0cm 13cm 0cm 2cm, clip=true,scale=0.4]{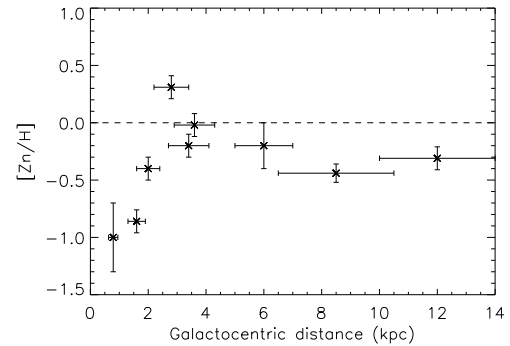}
\caption{[Zn/H] against Galactocentric distance of sources.}\label{GC_FeH}
\end{figure}

\begin{figure}
\includegraphics[trim=0cm 13cm 0cm 2cm, clip=true,scale=0.4]{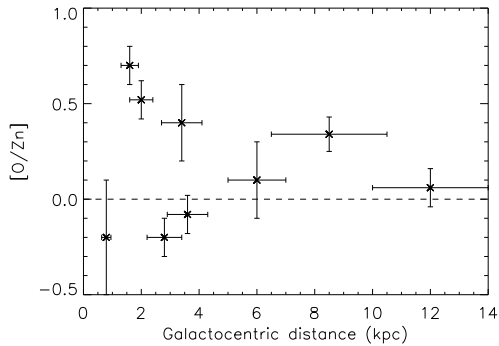}
\caption{[O/Zn] against Galactocentric distance of sources.}\label{GC_OFe}
\end{figure}

Fig. \ref{GC_FeH} and \ref{GC_OFe} show [Zn/H] and [O/Zn] as a function of the Galactocentric distances of the observed sources. 

Abundance gradients throughout the Milky Way have been previously studied. For example, \citet{Rich1998} shows a decline in [Fe/H] from $\sim0$ to $\sim-0.4$ over 1000pc in the Galactic bulge.  Results derived specifically from planetary nebulae generally focus on the gradients of O and Ne through the galaxy. \citet{Pottasch2006}, for example, find that the abundance of oxygen decreases over the range 3-11 kpc at a rate of $-0.085$ dex/kpc. \citet{Gutenkunst2009} derive abundance gradients for a sample of Galactic Bulge and Galactic Disk nebulae for Ne, S, Ar and O. They report negative gradients for the abundance trends of Galactic disk nebulae of between $-0.08$ and $-0.14$ dex/kpc. For Galactic bulge planetary nebulae, the results are less clear, with fits to the abundances having very large uncertainties. It is clear that the bulge and disk do not share an abundance trend with Galactocentric distance.

As shown in Fig. \ref{GC_FeH}, from our sample, [Zn/H] appears to increase with increasing R$_{\rm{G}}$ until $\sim4$ kpc, in contrast to \citet{Rich1998}  but more in agreement with the results of \citet{Gutenkunst2009} . After this point, there appears to be a shallow decline outwards,  in agreement with the results from \citet{Pottasch2006} and \citet{Gutenkunst2009} . However, due to the small sample size, this must be taken with caution.

Fig. \ref{GC_OFe} shows that around half of our sample lie within one sigma of Solar [O/Zn], with the rest showing enhancement in oxygen. No trend of [O/Zn] with Galactocentric distance is discernible.

 \section{Conclusions}

We have derived [Zn/H] and [O/Zn] for a sample of nine planetary nebulae: seven newly observed nebulae (five from the Galactic bulge and two from the Galactic disk) and two which were previously observed by \citet{Dinerstein2001}. 

Four out of five of those nebulae present in the Galactic bulge show sub-solar [Zn/H] and elevated [O/Zn]. The remaining bulge nebula shows an unusually high [Zn/H] and low [O/Zn]. 

One of the two newly observed Galactic disk nebulae, PNG 019.7+03.2, shows solar abundances for both [Zn/H] and [O/Zn], whereas the other, PNG 040.4--03.1, shows sub-solar [Zn/H] and elevated [O/Zn]. The latter, however, should be taken with caution as values for ionic abundances and nebular temperatures and densities  could not be computed as sufficient optical spectra were not available in the literature.

The two nebulae, NGC 7027 and IC 5117, as observed by \citet{Dinerstein2001} show [Zn/H] lower than Solar by more than a factor of 2. NGC 7027 also exhibits an enhanced [O/Zn] ratio in comparison to Solar, whereas IC 5117 displays [O/Zn] which is in line with Solar. The distinct difference from Solar values in NGC 7027 may be attributable to the large size of the nebula (14 arcseconds) resulting in a significant difference between the slit losses of the optical observations, from which the nebular temperature, densities and oxygen abundances were derived, and of the infra-red observations, from which the zinc abundances were derived.

Overall, we have found the majority of this sample to exhibit sub-solar [Zn/H], with the lowest being 10\% that of Solar. Half of the sample show [O/Zn] in line with the Solar value while the remaining half show enhancement in [O/Zn]. This will be further explored in a future paper that will present [Zn IV] observations of a larger sample of planetary nebulae from a wider range of Milky Way populations, and provide a broader context for our results on the Galactic bulge nebulae.

\section{Acknowledgements}

We thank Dr. Keith Butler for providing information on the [Zn IV] line emissivity prior to publication and the ESO staff at Paranal for support whilst observing. This research was supported in part by NSF grant AST 0708245 to H.L.D. We also acknowledge the support of the UK's Science and Technology Facilities Council. Based on observations made with ESO telescopes at the Paranal Observatory under programme ID 089.D-0084(A).

\bibliography{Zn_refs}
\end{document}